\documentclass[prd,aps,floatfix,nofootinbib,11 pt]{revtex4}
\usepackage{amssymb}
\usepackage{amsmath,graphicx,color,epsfig}

\input{tcilatex}

\begin{document}

\title{Quantum Parrondo's games under decoherence}
\author{Salman Khan\thanks{%
sksafi@phys.qau.edu.pk}, M. Ramzan\thanks{%
mramzan@phys.qau.edu.pk} and M. K. Khan}
\address{Department of Physics Quaid-i-Azam University \\
Islamabad 45320, Pakistan}
\date{\today }

\begin{abstract}
We study the effect of quantum noise on history dependent quantum Parrondo's
games by taking into account different noise channels. Our calculations show
that entanglement can play a crucial role in quantum Parrondo's games. It is
seen that for the maximally entangled initial state in the presence of
decoherence, the quantum phases strongly influence the payoffs for various
sequences of the game. The effect of amplitude damping channel leads to
winning payoffs. Whereas the depolarizing and phase damping channels lead to
the losing payoffs. In case of amplitude damping channel, the payoffs are
enhanced in the presence of decoherence for the sequence $AAB$. This is
because the quantum phases interfere constructively which leads to the
quantum enhancement of the payoffs in comparison to the undecohered case. It
is also seen that the quantum phase angles damp the payoffs significantly in
the presence of decoherence. Furthermore, it is seen that for multiple games
of sequence $AAB$, under the influence of amplitude damping channel, the
game still remains a winning game. However, the quantum enhancement reduces
in comparison to the single game of sequence $AAB$ because of the
destructive interference of phase dependent terms. In case of depolarizing
channel, the game becomes a loosing game. It is seen that for the game
sequence $B$ the game is loosing one and the behavior of sequences $B$ and $%
BB$ is similar for amplitude damping and depolarizing channels. In addition,
the repeated games of $A$ are only influenced by the amplitude damping
channel and the game remains a losing game. Furthermore, it is also seen
that for any sequence when played in series, the phase damping channel does
not influence the game.\newline
\end{abstract}

\pacs{02.50.Le; 03.65.Ud; 03.67.-a}
\maketitle

\address{Department of Physics Quaid-i-Azam University \\
Islamabad 45320, Pakistan}

Keywords: Quantum Parrondo's games; decoherence; payoffs\newline

\vspace*{1.0cm}

\vspace*{1.0cm}


\section{Introduction}

Game theory \cite{Rasmusen} has been implemented for diverse applications in
different areas for example; economics, evolutionary biology, psychology and
physics. More recently, the game theory is being used to model distributed
and parallel computing in the field of computer science. It is the theory of
decision making and conflict between different agents.\ Starting from the
works of Meyer and Eisert, quantum game theory has been recognized as an
important theory with useful applications [2-11].

In quantum game theory, the initial state entanglement has produced
interesting results. Quantum entanglement is one of the fascinating features
of quantum mechanics and plays a crucial role in quantum information
processing as well. When quantum information processing is performed in the
real world, the decoherence caused by an external environment is inevitable.
Decoherence effects in different quantum games have been studied in refs.
\cite{Flitney1,Chen,Hollen}. Here in this work, we are interested to study
the decoherence effects on the history dependent quantum Parrondo's games
played in various sequences.

In the context of classical Parrondo's games, the two games that are losing
when played individually can be combined in various sequences to produce a
winning game \cite{Har,Har1}. Parrondo's games have attracted considerable
attention in the past as they can be related to physical systems such as the
Brownian ratchet \cite{Har3}, lattice gas automata \cite{Meyer} and spin
systems \cite{Moraal}. Based on the maximal entanglement between the qubits,
a quantization protocol for the history dependent Parrondo's games was
proposed by Flitney et al. \cite{Flitney}. Multi-player extension to
classical Parrondo's games was given by Toral \cite{Toral}.

In this paper, we study the effect of decoherence on history dependent
quantum Parrondo's games by considering different prototype channels such as
amplitude damping, depolarizing and phase damping channels, parameterized by
the decoherence parameter $p\in \lbrack 0,1]$. The lower and upper limits of
the decoherence parameter $p$ correspond to a fully coherent and fully
decohered systems, respectively. We study the effect of quantum decoherence
on the game dynamics. It is seen that the payoffs are enhanced due to the
presence of decoherence in case of amplitude damping channel for the single
game of sequence $AAB$. The enhancement in payoffs occurs due to the
constructive interference of quantum phases, $\delta $ and $\beta _{i}$. It
is also seen that the increase in payoffs for phase damping and depolarizing
channels is not much prominent in comparison to the amplitude damping
channel. We also analyze the influence of decoherence by playing other
sequences such as $B$, $BB$, $BBB$, $AA...A$ and multiple games of sequence $%
AAB$. The results are discussed in detail in the results and discussions
section.

\section{History dependent quantum Parrondo's games}

History dependent Parrondo's games consist of two simpler coin tossing
games; $A$ and $B$. Game $A$ is straight forward one player biased coin
flipping game that wins $1$ when lands head up and loses $1$ when it lands
tail up. However, game $B$ consists of four biased coins, the selection of
each of them depends on history of the games (the results of previous two
games). In the classical version, the winning probabilities for coin $A$ and
for each coin of game $B$ are given by%
\begin{equation}
p_{0}=\frac{1}{2}-\epsilon ,\quad p_{1}=\frac{7}{10}-\epsilon ,\quad
p_{2}=p_{3}=\frac{1}{4}-\epsilon ,\quad p_{4}=\frac{9}{10}-\epsilon
\label{1}
\end{equation}%
respectively. It is shown in ref. \cite{Har3} that for a small positive
value of $\epsilon $ each of the games $A$ and $B$ played individually is a
losing game, however, if they are played in various sequences of $A$ and $B$
produce a winning result.

The game $A$\ can be quantized by replacing the tossing of a coin by an $%
SU(2)$ operator on a qubit as given in ref. \cite{Flitney}%
\begin{equation}
A(\theta ,\gamma ,\delta )=\left[
\begin{array}{cc}
\exp \left( \frac{-i(\gamma +\delta )}{2}\right) \cos \theta & -\exp \left(
\frac{-i(\gamma -\delta )}{2}\right) \sin \theta \\
\exp \left( \frac{i(\gamma -\delta )}{2}\right) \sin \theta & \exp \left(
\frac{i(\gamma +\delta )}{2}\right) \cos \theta%
\end{array}%
\right]  \label{2}
\end{equation}%
where $\theta \in \left[ -\pi ,\text{ }\pi \right] ,$ $\gamma $ and $\delta
\in \left[ 0,\text{ }2\pi \right] .$ Similarly, the operator for game $B$
consists of four $SU(2)$ operations, each of the form given in equation (\ref%
{2}), where the choice of the use of these operations depends on the outcome
of the previous two games:%
\begin{equation}
B=\left[
\begin{array}{cccc}
A_{1} & 0 & 0 & 0 \\
0 & A_{2} & 0 & 0 \\
0 & 0 & A_{3} & 0 \\
0 & 0 & 0 & A_{4}%
\end{array}%
\right]  \label{3}
\end{equation}%
with $A_{i}=A\left( \phi _{i},\alpha _{i},\beta _{i}\right) $. The operator $%
B$ acts on the following three qubits state

\begin{equation}
|\Psi _{2}\rangle \otimes |\Psi _{1}\rangle \otimes |\Psi _{0}\rangle
\label{E3}
\end{equation}%
where $|\Psi _{2}\rangle $ and $|\Psi _{1}\rangle $ stand for the results of
two successive previous games and $|\Psi _{0}\rangle $ stands for the target
qubit that results in some output, say $b$. Each qubit in equation (\ref{E3}%
) could be in one of the possible states of a qubit. The final density
matrix of the game is given by%
\begin{equation}
\rho _{f}=U\rho _{i}U^{\dag }  \label{4}
\end{equation}%
For $n$ successive games of sequence $B,$ the operator $U$\ can be written as

\begin{eqnarray}
U_{B} &=&\left( I^{\otimes n-1}\otimes B\right) \left( I^{\otimes
n-2}\otimes B\otimes I\right) \left( I^{\otimes n-3}\otimes B\otimes
I^{\otimes 2}\right)  \label{5} \\
&&...\left( I\otimes B\otimes I^{\otimes n-2}\right) \left( B\otimes
I^{\otimes n-1}\right)  \notag
\end{eqnarray}%
where $I$ represents the single qubit identity operator. Similarly, the
operator $U$ for $n$ games of sequence $AAB$ can be written as

\begin{eqnarray}
U_{AAB}^{n} &=&\left( I^{\otimes 3n-3}\otimes \left( B\left( A\otimes
A\otimes I\right) \right) \right) \left( I^{\otimes 3n-6}\otimes \left(
B\left( A\otimes A\otimes I\right) \right) \otimes I^{\otimes 3}\right)
\notag \\
&&...\left( \left( B\left( A\otimes A\otimes I\right) \right) \otimes
I^{\otimes 3n-3}\right)  \notag \\
&=&U^{\otimes n}  \label{7}
\end{eqnarray}%
where $U=B\left( A\otimes A\otimes I\right) $ stands for the operator of a
single game sequence $AAB$. Here in this paper, we consider various
sequences of Parrondo's games such as $AA...A$, $B$, $BB$, $BBB$, $AAB$ and
a series of $AAB$. To study the effect of decoherence, we restrict our
calculations only to the maximally entangled initial state of the form

\begin{equation}
|\Psi _{i}\rangle =\frac{1}{\sqrt{2}}\left( |00...0\rangle +|11...1\rangle
\right)  \label{8}
\end{equation}%
We consider $|0\rangle $ as "loss" state and the $|1\rangle $ as a "win"
state. Furthermore, we can fix the computational basis of the Hilbert Space
for example $\mathcal{H}^{\otimes 3}$ for a single game sequence $AAB$ in
the basis ordered as $|000\rangle ,$ $|001\rangle ,$ $|010\rangle ,$ $%
|011\rangle ,$ $|100\rangle ,$ $|101\rangle ,$ $|110\rangle $ and $%
|111\rangle ,$ respectively.

\section{Quantum channels}

A natural way to describe the dynamics of a quantum system is to consider it
as arising from an interaction between the system and the environment. In
general quantum systems are prone to decoherence effects and it is important
to analyze these effects in real practical situations. Environmental
interactions can destroy the important features of quantum computation.
However, quantum error correction \cite{Lidar} and decoherence free
subspaces \cite{Presk} can be used to perform quantum computing even in the
presence of noise. In the most general case, the quantum evolution can be
described by the superoperator $\Phi $, which can be expressed in Kraus
operator representation as \cite{Nielson}
\begin{equation}
\Phi \left( \rho \right) =\sum_{k}E_{k}\rho E_{k}^{\dag }  \label{9}
\end{equation}%
where%
\begin{equation}
\sum_{k}E_{k}^{\dag }E_{k}=I  \label{10}
\end{equation}%
Superoperators provide a way to describe the evolution of quantum states in
a noisy environment. In our scheme, the Kraus operators are of the dimension
$2^{3}$. They are constructed from one qubit operators by taking their
tensor product over all $n^{3}$ combinations of $\pi (i)$ indices
\begin{equation}
E_{k}=\dbinom{{\LARGE \otimes }}{\pi }e_{\pi (i)}  \label{11}
\end{equation}%
where $n$ is the number of Kraus operators for a single qubit channel. The
single qubit Kraus operators for quantum channels considered in this paper
are given in table 1.

\section{Results and discussions}

In this section we present our results for different sequences of Parrondo's
games in the presence of various noisy channels. The final density matrix of
the game after the action of a channel (as specified in equation \ref{9})
for example, for a single game sequence $AAB$, can be written as
\begin{equation}
\rho ^{AAB}=\Phi \rho _{i}^{AAB}  \label{12}
\end{equation}%
where $\rho _{i}^{AAB}=|\Psi ^{AAB}\rangle \langle \Psi ^{AAB}|$ and $|\Psi
^{AAB}\rangle =\frac{1}{\sqrt{2}}\left( |000\rangle +|111\rangle \right) $.
The game's final density matrix after the application of operator $U$ can be
computed by using equation (\ref{4}). To determine the payoff, we assume
that the payoff for a $|1\rangle $\ state is $+1$, and is $-1$ for a $%
|0\rangle $ state. The total payoff can be determined by using the relation%
\begin{equation}
\langle \$\rangle =\underset{ijk}{\sum }(l+m+n)\rho _{ijk}  \label{14}
\end{equation}%
where $l=(-1)^{i+1},\;m=(-1)^{j+1},\;n=(-1)^{k+1},$ the indices $i,j$ and $k$
run from $0$ to $1$ and $\rho _{ijk}$ represent the diagonal elements of the
final density matrix.

The payoff for amplitude damping channel for the game sequence $AAB$ can be
written as

\begin{eqnarray}
\langle \$^{\text{AD}}\rangle &=&3p+\cos ^{2}\phi _{1}-\cos ^{2}\phi
_{4}+p[\{-4+(5-2p)p\}\cos ^{2}\phi _{1}  \notag \\
&&+(-1+2p)\{(-1+p)(\cos ^{2}\phi _{2}+\cos ^{2}\phi _{3})-p\cos ^{2}\phi
_{4}\}]  \notag \\
&&+\cos ^{4}\theta \{p(-3+6p-4p^{2})\cos 2\phi _{1}+p(3-6p+4p^{2})\cos 2\phi
_{2}  \notag \\
&&+2p(3-6p+4p^{2})\cos ^{2}\phi _{3}-2p(3-6p+4p^{2})\cos ^{2}\phi _{4}
\notag \\
&&+(1-p)^{3/2}(-\cos (2\delta +\beta _{1})\sin 2\phi _{1}+\cos (2\delta
+\beta _{2})\sin 2\phi _{2}  \notag \\
&&+\cos (2\delta +\beta _{3})\sin 2\phi _{3}-\cos (2\delta +\beta _{4})\sin
2\phi _{4})\}  \notag \\
&&+\cos ^{2}\theta \lbrack -4p+2(1-2p)^{2}(-1+p)\cos ^{2}\phi
_{1}-2p(3-6p+4p^{2})\cos ^{2}\phi _{2}  \notag \\
&&-2p(3-6p+4p^{2})\cos ^{2}\phi _{3}+2\cos ^{2}\phi _{4}+2(1-2p)^{2}p\cos
^{2}\phi _{4}  \notag \\
&&+\sqrt{1-p}\cos (2\delta +\beta _{1})\sin 2\phi _{1}+\sqrt{1-p}\{-p\cos
(2\delta +\beta _{1})\sin 2\phi _{1}+(-1+  \notag \\
&&p)(\cos (2\delta +\beta _{2})\sin 2\phi _{2}+\cos (2\delta +\beta
_{3})\sin 2\phi _{3}-\cos (2\delta +\beta _{4})\sin 2\phi _{4})\}]  \notag \\
&&  \label{E15}
\end{eqnarray}%
The payoff obtained for the case of depolarizing channel is given as
\begin{eqnarray}
\langle \$^{\text{DP}}\rangle &=&\left( -1+p\right) ^{2}(2\cos 2\theta
\left( -\cos ^{2}\phi _{1}+\cos ^{2}\phi _{4}\right) +\left( -1+p\right)
\cos ^{2}\theta \sin ^{2}\theta  \notag \\
&&(-\cos \left( 2\delta +\beta _{1}\right) \sin 2\phi _{1}+\cos \left(
2\delta +\beta _{2}\right) \sin 2\phi _{2}  \notag \\
&&+\cos \left( 2\delta +\beta _{3}\right) \sin 2\phi _{3}-2\cos \left(
2\delta +\beta _{4}\right) \sin 2\phi _{4})  \label{16}
\end{eqnarray}%
The payoff in case of phase damping channel becomes%
\begin{eqnarray}
\langle \$^{\text{PD}}\rangle &=&\cos 2\theta \left( -\cos ^{2}\phi
_{1}+\cos ^{2}\phi _{4}\right)  \notag \\
&&-\left( 1-p\right) ^{3/2}\cos ^{2}\theta \sin ^{2}\theta (-\cos \left(
2\delta +\beta _{1}\right) \sin 2\phi _{1}  \notag \\
&&+\cos \left( 2\delta +\beta _{2}\right) \sin 2\phi _{2}+\cos \left(
2\delta +\beta _{3}\right) \sin 2\phi _{3}  \notag \\
&&-2\cos \left( 2\delta +\beta _{4}\right) \sin 2\phi _{4})  \label{17}
\end{eqnarray}%
where the superscripts AD, DP and PD represent the amplitude damping,
depolarizing and phase damping channels, respectively. The winning
probability for a flip of a coin is equal to the square of the the sine of
the rotation angle (i.e. $\theta $ and $\phi _{i}$). Whereas $\delta $ and $%
\beta _{i}$'s\ represent the quantum phases. It can be easily checked that
the result of ref. \cite{Flitney} for maximally entangled state can be
reproduced by setting the decoherence parameter $p=0$\ in equations (\ref%
{E15}, \ref{16} and \ref{17}). The presence of phase angles $\delta $ and $%
\beta _{i}$'s in the equations (\ref{E15}, \ref{16} and \ref{17}) leads to a
range of payoffs for the set of values of $\theta $ and $\phi _{i}$'s
corresponding to the classical winning probabilities of the two games as
given in equation (\ref{1}).

In figure 1, we have plotted the payoffs for the single sequence of game $%
AAB $ as a function of decoherence parameter, $p$, for different noise
channels. The rotation angles correspond to the classical winning
probabilities and the quantum phase angles are taken as$\ \beta _{1}=\beta
_{2}=\pi /2,$ $\beta _{3}=\pi /4,$ $\beta _{4}=\pi /3,$ and $\delta =\pi /5.$
It is seen that the payoff of the game is further enhanced under decoherence
for the amplitude damping channel. The significant increase in payoff for
amplitude damping channel results from the constructive interference arising
due to the presence of decoherence and the quantum phases. However, under
the action of depolarizing and phase damping channels, destructive
interference is seen and the game's sequence $AAB$ becomes a losing one. In
figure 2, the payoffs as a function of quantum phase angle $\delta $ are
plotted for all the three channels. It is seen that for a particular choice
of values of the parameters, the payoffs become positive only for the range $%
\frac{\pi }{4}$ $\lesssim \delta \lesssim \frac{3\pi }{4}$.and varies
symmetrically around $\frac{\pi }{2}$. The payoffs are also significantly
damped in comparison to the undecohered case. In figure 3 we have plotted
the payoffs as a function of quantum phase angle $\beta _{1}$. It is seen
that the game becomes a winning game. The effect of quantum phase angle $%
\beta _{2}$ on the payoffs is shown in figure 4. In figures 5 and 6, we have
plotted the payoffs versus the phase angles $\beta _{3}$ and\ $\beta _{4}$
respectively\textbf{.} It is shown that the game behaves as a loosing and
winning one respectively. Similar to the behaviour of the game against the
phase angle $\delta ,$\ the payoffs are significantly damped against the
phase angles $\beta _{i}$'s under decoherence.

In figure 7, we have plotted the expected payoffs as a function of
decoherence under the influence of various noisy channels by playing the
sequence $AAB$ repeatedly at $\varepsilon =\frac{1}{168}$, where the inset
figure corresponds to $\varepsilon =\frac{1}{112}$. The expected payoffs for
this generalized case for different channels are obtained as%
\begin{eqnarray}
\langle \$^{\text{AD}}\rangle &=&(\frac{1}{60}p+(\frac{2}{15}%
-2.27p+0.27p^{2})\varepsilon )  \notag \\
\langle \$^{\text{DP}}\rangle &=&(\frac{2}{15}-0.35p+0.24p^{2})\varepsilon
\notag \\
\langle \$^{\text{PD}}\rangle &=&\frac{2}{15}\varepsilon  \label{18}
\end{eqnarray}%
The results given in equation (\ref{18}) are obtained at the maximum payoffs
condition (i.e. $\beta _{2}=\beta _{3}=\pi -2\delta $ and $\beta _{1}=\beta
_{4}=-2\delta $) as given in ref. \cite{Flitney}. From figure 7, one can see
that for $\varepsilon =\frac{1}{168}$, the amplitude damping channel leads
to the winning game. It is also seen that the quantum enhancement reduces
when we play a repeated sequence of $AAB$ on the maximally entangled initial
state, because of the destructive interference of phase dependent terms.
Whereas the depolarizing channel corresponds to the loosing game. However,
from the inset figure, it is seen that in the case of amplitude damping
channel for the values of decoherence parameter $p$\ in the range $0.5<p<1,$
the repeated sequence $AAB$ leads to a negative payoff. Furthermore, it is
clear from equation (\ref{18}) that the phase damping channel does not
influence the game. The results for a single sequence of game $B$ under the
action of three channels are obtained as%
\begin{eqnarray}
\langle \$^{\text{AD}}\rangle &=&\frac{1}{30}[2+p\{-7-20\varepsilon
+p(-29+22p)\}]  \notag \\
\langle \$^{\text{DP}}\rangle &=&\frac{1}{135}(3-4p)^{2}  \notag \\
\langle \$^{\text{PD}}\rangle &=&\frac{1}{15}  \label{19}
\end{eqnarray}%
In case of sequence $BB$, the payoffs become

\begin{eqnarray}
\langle \$^{\text{AD}}\rangle &=&\frac{1}{400}[13-2p\{67+2p(51-64p+22p^{2})\}
\notag \\
&&+10\varepsilon \{2+p(-11-62p+44p^{2})\}]  \notag \\
\langle \$^{\text{DP}}\rangle &=&\frac{1}{32400}[(3-4p)^{2}\{117+180%
\varepsilon +88(3-2p)p\}]  \notag \\
\langle \$^{\text{PD}}\rangle &=&\frac{13}{400}+\frac{\varepsilon }{20}
\label{20}
\end{eqnarray}%
Similarly, the payoffs for $BBB$ sequence of the game can be written as%
\begin{eqnarray}
\langle \$^{\text{AD}}\rangle &=&(0.017-0.41p-0.13p^{2}+0.45p^{3})  \notag \\
&&+(0.03-1.11p+0.44p^{2}-2.66p^{3})\varepsilon  \notag \\
\langle \$^{\text{DP}}\rangle &=&(0.017+0.01p-0.13p^{2}+0.10p^{3})  \notag \\
&&+(0.03+0.15p-0.73p^{2}+0.83p^{3})\varepsilon  \notag \\
\langle \$^{\text{PD}}\rangle &=&0.017+0.03\varepsilon  \label{21}
\end{eqnarray}%
One can easily check that by setting $p=0$, in equations (\ref{18}-\ref{21}%
), the results of ref. \cite{Flitney} can be reproduced. In figures 8 and 9,
we have plotted the expected payoffs as a function of the decoherence
parameter, $p$ for $\varepsilon =\frac{1}{168}$ for $B$'s game sequences $BB$
and $BBB$\textbf{\ }respectively\textbf{. }It can be inferred from the
results for\ game $B$'s sequences that the behavior of sequences $B$ and $BB$
is similar for amplitude damping and depolarizing channels. It is seen that
the games are losing ones for both the channels. However, for higher values
of decoherence parameter $p$, the payoff starts increasing in case of
depolarizing channel. We have also calculated the results for the series of
game $A$, the payoff under the influence of amplitude damping channel
becomes $-\frac{3}{32}\varepsilon p$. As both $\varepsilon $ and $p$ are
positive, the negative sign ensures that for any value of decoherence
parameter $p$, the series of game $A$ remains a losing one. However, the
payoffs obtained for the series of game $A$ under the action of depolarizing
and phase damping channels are zero.

\section{Conclusion}

We study the history dependent quantum Parrondo's games under the effect of
decoherence being played in different sequences for amplitude damping,
depolarizing and phase damping channels. It is seen that the payoffs are
enhanced in the presence decoherence for maximally entangled state in case
of a single game of the sequence $AAB$ for amplitude damping channel. The
decoherence causes constructive interference of quantum phases that leads to
the enhancement of payoff in comparison to the undecohered case. Whereas in
case of the depolarizing and the phase damping channels, phases interfere
destructively that results into a decrease in payoffs (a losing game
sequence). It is also seen that the payoffs are significantly damped against
the quantum phases in the presence of decoherence.

Furthermore, it is seen that for repeated games of sequence $AAB$, under the
influence of amplitude damping channel the game becomes a winning game.
However, the quantum enhancement in payoff reduces in this case as compared
to the single game of sequence $AAB$. Whereas under the influence of
depolarizing channel the game becomes a loosing game. It is shown that the
games $B$ and $BB$ behave similarly for amplitude damping and depolarizing
channels. It is seen that the repeated games of $A$ are only influenced by
the amplitude damping channel and it always remain a losing game.
Furthermore, it is also seen that every sequence played repeatedly remains
unaffected under the influence of phase damping channel.

\textbf{Acknowledgement}

One of the authors (Salman Khan) is thankful to World Federation of
Scientists for partial financial support under the National Scholarship
Program for Pakistan.

{\huge Figures Captions}\newline
\textbf{Figure 1.} The expected payoffs for a single game of sequence $AAB$
are plotted as a function of the decoherence parameter, $p$ for amplitude
damping channel (AD), depolarizing channel (DP), phase damping channel (PD)
and Flitney and Abbott results (FNA) with $\delta =\frac{\pi }{5}$, $\beta
_{1}=\frac{\pi }{2}$, $\beta _{2}=\frac{\pi }{2}$, $\beta _{3}=\frac{\pi }{4}
$, $\beta _{4}=\frac{\pi }{3}$ and $\varepsilon =\frac{1}{168}$.\newline
\textbf{Figure 2.} The expected payoffs for a single game of sequence $AAB$
are plotted as a function of the phase angle, $\delta $ for amplitude
damping channel, depolarizing channel, phase damping channel and FNA results
with $p=0.5$, $\beta _{1}=\frac{\pi }{2}$, $\beta _{2}=\frac{\pi }{3}$, $%
\beta _{3}=\frac{\pi }{4}$, $\beta _{4}=\frac{\pi }{3}$ and $\varepsilon =%
\frac{1}{168}$.\newline
\textbf{Figure 3. }The expected payoffs for a single game of sequence $AAB$
are plotted as a function of the quantum phase angle, $\beta _{1}$ for
amplitude damping channel, depolarizing channel, phase damping channel and
FNA results with $p=0.5$, $\delta =\frac{\pi }{2}$, $\beta _{2}=\frac{\pi }{3%
}$, $\beta _{3}=\frac{\pi }{2}$, $\beta _{4}=\pi $ and $\varepsilon =\frac{1%
}{168}$ respectively.\newline
\textbf{Figure 4.} The expected payoffs for a single game of sequence $AAB$
are plotted as a function of the quantum phase angle, $\beta _{2}$ for
amplitude damping channel, depolarizing channel, phase damping channel and
FNA results with $p=0.5$, $\delta =\pi $, $\beta _{1}=\frac{\pi }{2}$, $%
\beta _{3}=\pi $, $\beta _{4}=\frac{\pi }{2}$ and $\varepsilon =\frac{1}{168}
$.\newline
\textbf{Figure 5.} The expected payoffs for a single game of sequence $AAB$
are plotted as a function of the quantum phase angle, $\beta _{3}$ for
amplitude damping channel, depolarizing channel, phase damping channel and
FNA results with $p=0.5$, $\delta =\frac{\pi }{2}$, $\beta _{1}=2\pi $, $%
\beta _{2}=\frac{\pi }{6}$, $\beta _{4}=\pi $ and $\varepsilon =\frac{1}{168}
$.\newline
\textbf{Figure 6.} The expected payoffs for a single game of sequence $AAB$
are plotted as a function of the quantum phase angle, $\beta _{4}$ for
amplitude damping channel, depolarizing channel, phase damping channel and
FNA results with $p=0.5$, $\delta =\frac{\pi }{2}$, $\beta _{1}=\frac{\pi }{4%
}$, $\beta _{2}=\frac{\pi }{4}$, $\beta _{3}=\frac{\pi }{4}$ and $%
\varepsilon =\frac{1}{168}$.\newline
\textbf{Figure 7. }The expected payoffs for a series of sequence $AAB$ are
plotted as a function of the decoherence parameter, $p$ for amplitude
damping and depolarizing channels with $\varepsilon =\frac{1}{168}$ (for
inset figure, $\varepsilon =\frac{1}{112}$).\newline
\textbf{Figure 8.} The expected payoffs for the game sequence $BB$ are
plotted as a function of the decoherence parameter, $p$ for amplitude
damping and depolarizing channels with $\varepsilon =\frac{1}{112}$.\newline
\textbf{Figure 9.} The expected payoffs for the game sequence $BBB$ are
plotted as a function of the decoherence parameter, $p$ for amplitude
damping and depolarizing channels with $\varepsilon =\frac{1}{112}$.\newline
{\huge Table Caption}\newline
Table 1. Single qubit Kraus operators for typical noise channels such as
depolarizing, amplitude damping and phase damping channels where $p$
represents the decoherence parameter.{\huge \ }
\begin{figure}[tbp]
\begin{center}
\vspace{-2cm} \includegraphics[scale=0.6]{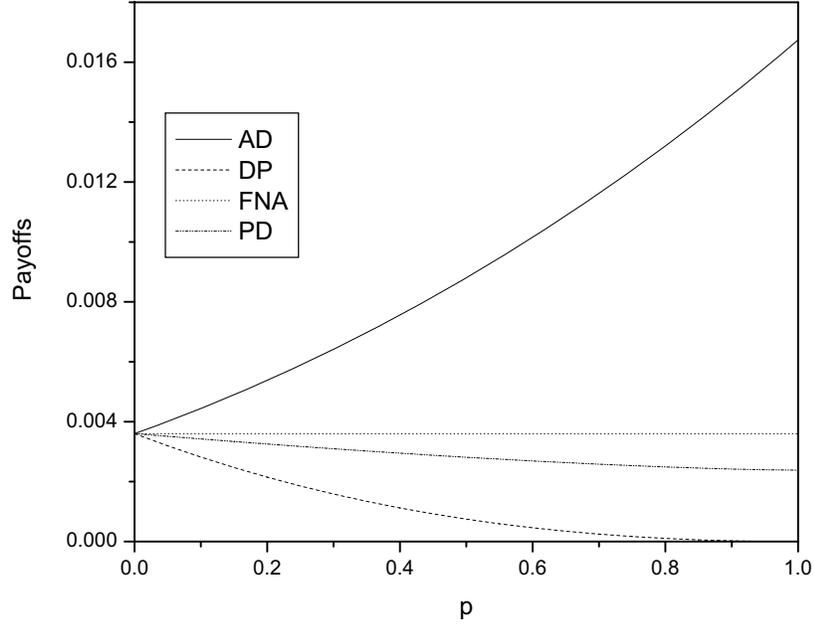} \\[0pt]
\end{center}
\caption{The expected payoffs for a single game of sequence $AAB$ are
plotted as a function of the decoherence parameter, $p$ for amplitude
damping channel (AD), depolarizing channel (DP), phase damping channel (PD)
and Flitney and Abbott results (FNA) with $\protect\delta =\frac{\protect\pi
}{5}$, $\protect\beta _{1}=\frac{\protect\pi }{2}$, $\protect\beta _{2}=%
\frac{\protect\pi }{2}$, $\protect\beta _{3}=\frac{\protect\pi }{4}$, $%
\protect\beta _{4}=\frac{\protect\pi }{3}$ and $\protect\varepsilon =\frac{1%
}{168}$.}
\end{figure}

\begin{figure}[tbp]
\begin{center}
\vspace{-2cm} \includegraphics[scale=0.6]{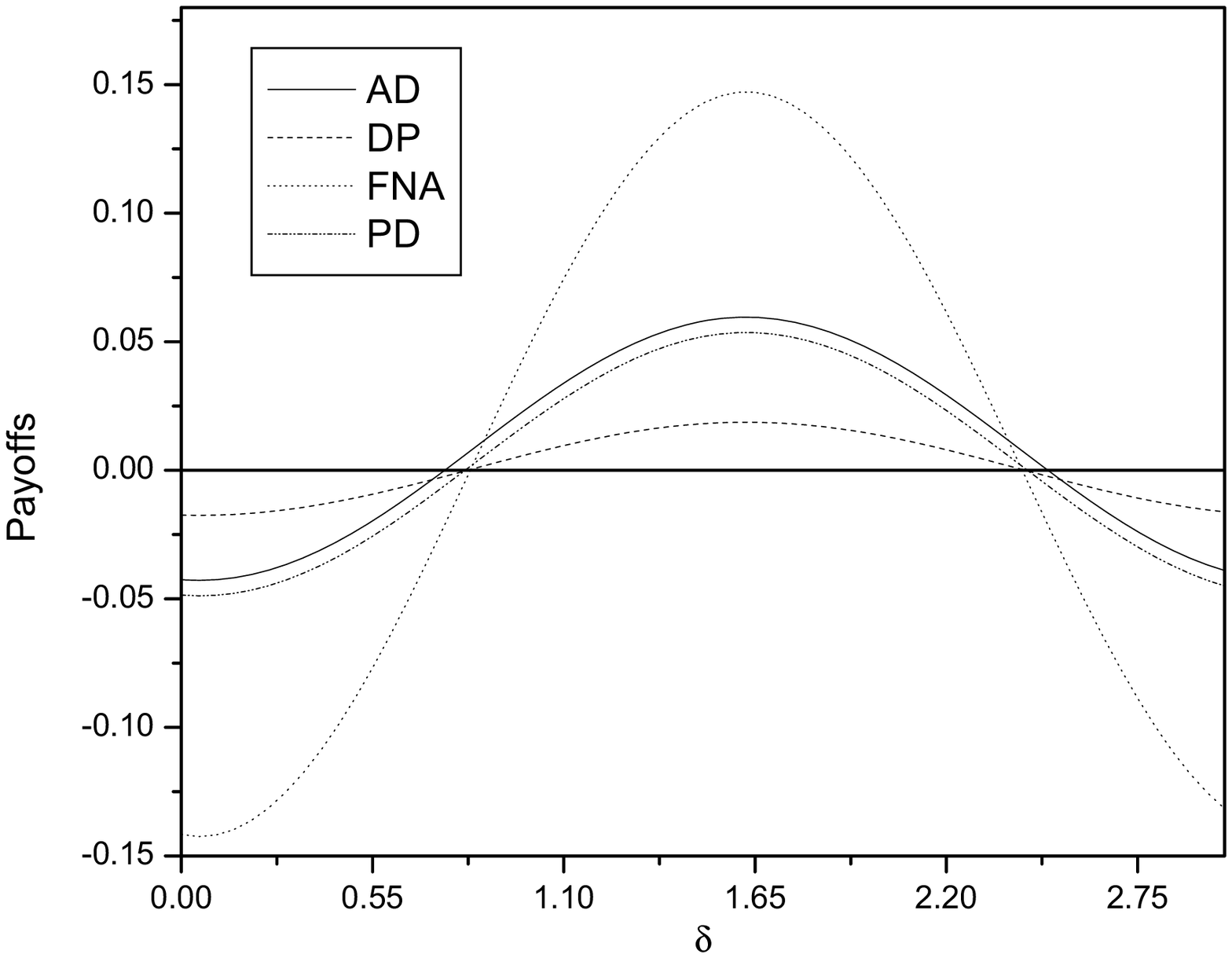} \\[0pt]
\end{center}
\caption{The expected payoffs for a single game of sequence $AAB$ are
plotted as a function of the phase angle, $\protect\delta $ for amplitude
damping channel, depolarizing channel, phase damping channel and FNA results
with $p=0.5$, $\protect\beta _{1}=\frac{\protect\pi }{2}$, $\protect\beta %
_{2}=\frac{\protect\pi }{3}$, $\protect\beta _{3}=\frac{\protect\pi }{4}$, $%
\protect\beta _{4}=\frac{\protect\pi }{3}$ and $\protect\varepsilon =\frac{1%
}{168}$.}
\end{figure}

\begin{figure}[tbp]
\begin{center}
\vspace{-2cm} \includegraphics[scale=0.6]{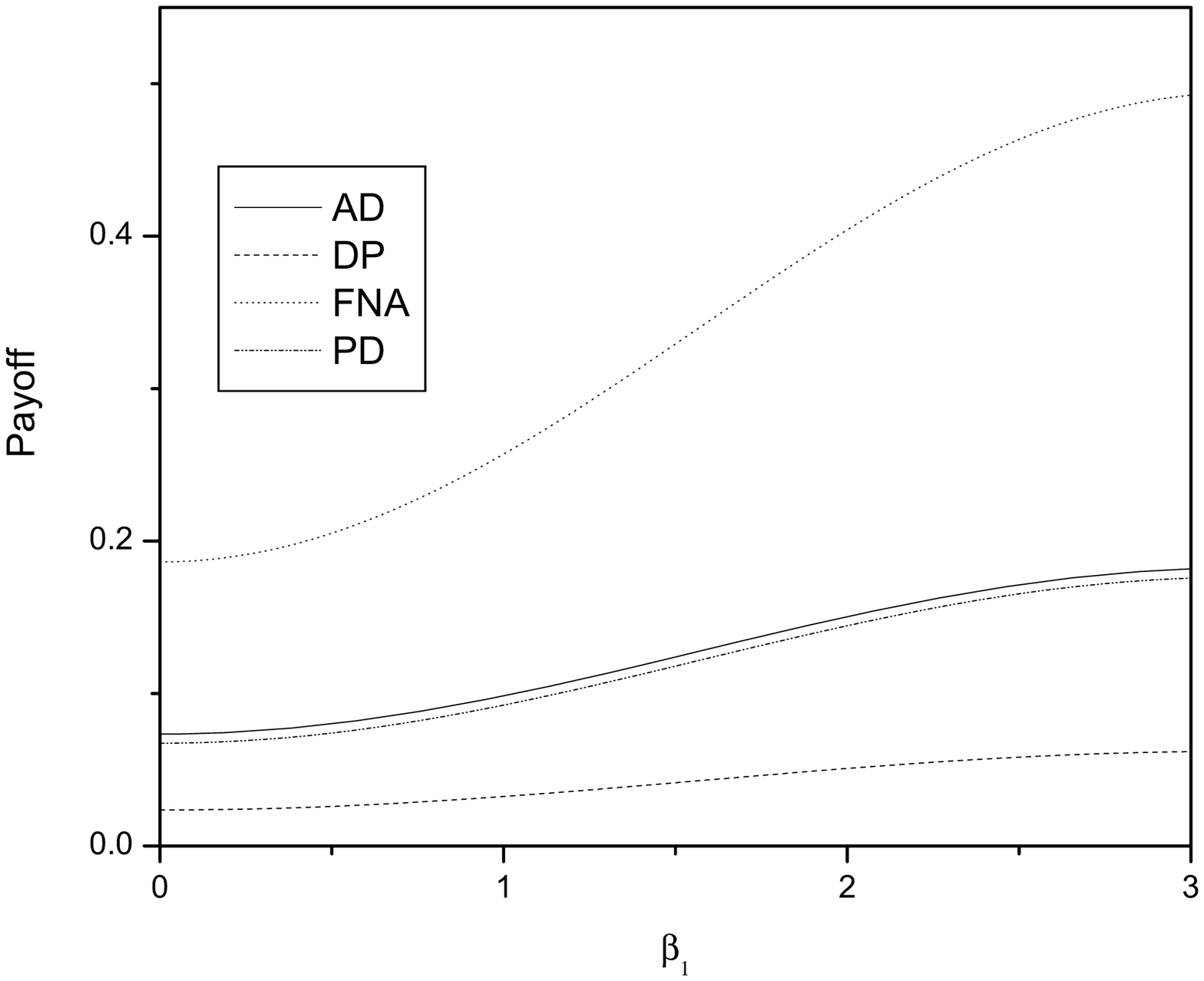} \\[0pt]
\end{center}
\caption{The expected payoffs for a single game of sequence $AAB$ are
plotted as a function of the quantum phase angle, $\protect\beta _{1}$ for
amplitude damping channel, depolarizing channel, phase damping channel and
FNA results with $p=0.5$, $\protect\delta =\frac{\protect\pi }{2}$, $\protect%
\beta _{2}=\frac{\protect\pi }{3}$, $\protect\beta _{3}=\frac{\protect\pi }{2%
}$, $\protect\beta _{4}=\protect\pi $ and $\protect\varepsilon =\frac{1}{168}
$ respectively.}
\end{figure}

\begin{figure}[tbp]
\begin{center}
\vspace{-2cm} \includegraphics[scale=0.6]{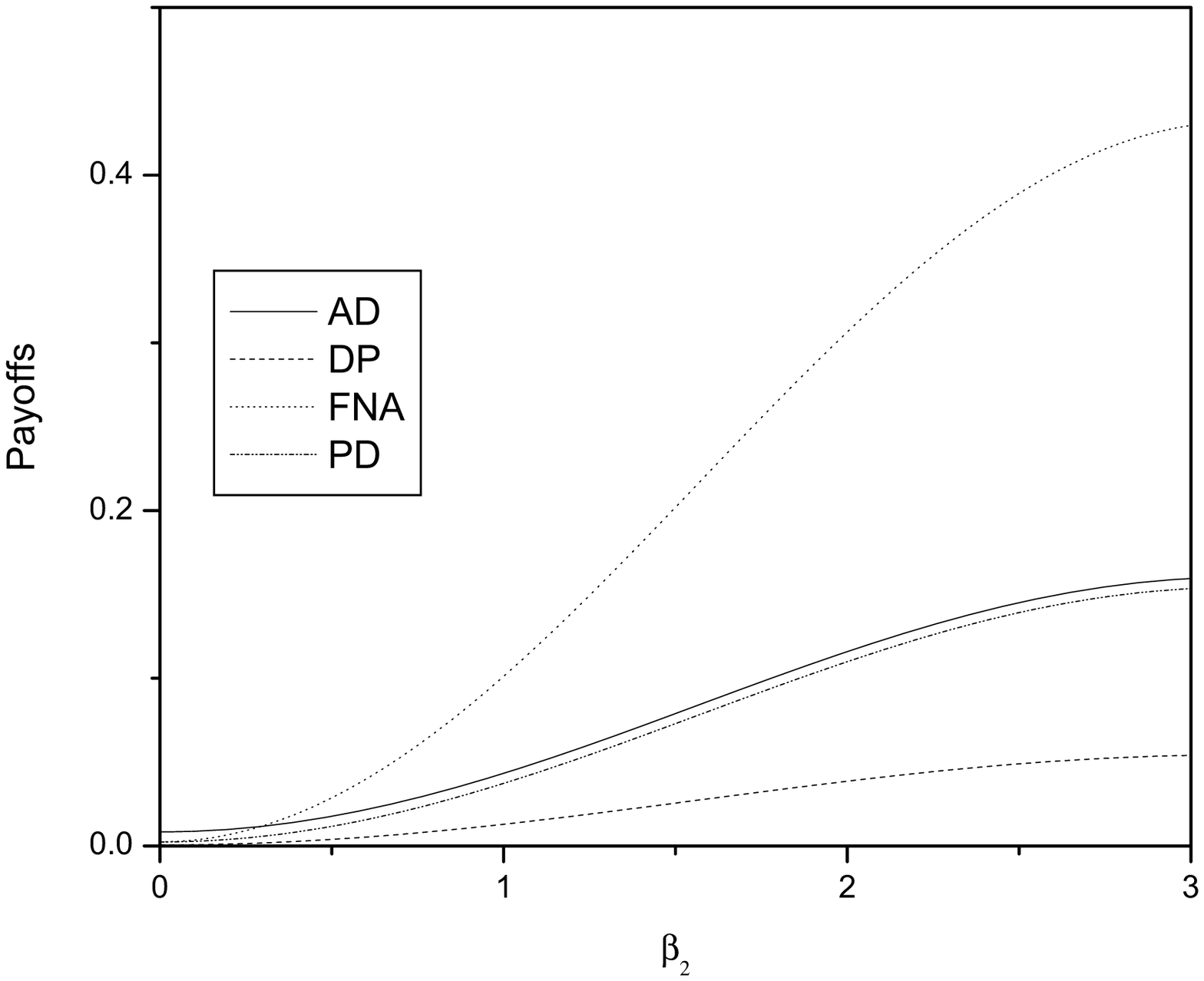} \\[0pt]
\end{center}
\caption{The expected payoffs for a single game of sequence $AAB$ are
plotted as a function of the quantum phase angle, $\protect\beta _{2}$ for
amplitude damping channel, depolarizing channel, phase damping channel and
FNA results with $p=0.5$, $\protect\delta =\protect\pi $, $\protect\beta %
_{1}=\frac{\protect\pi }{2}$, $\protect\beta _{3}=\protect\pi $, $\protect%
\beta _{4}=\frac{\protect\pi }{2}$ and $\protect\varepsilon =\frac{1}{168}$.}
\end{figure}

\begin{figure}[tbp]
\begin{center}
\vspace{-2cm} \includegraphics[scale=0.6]{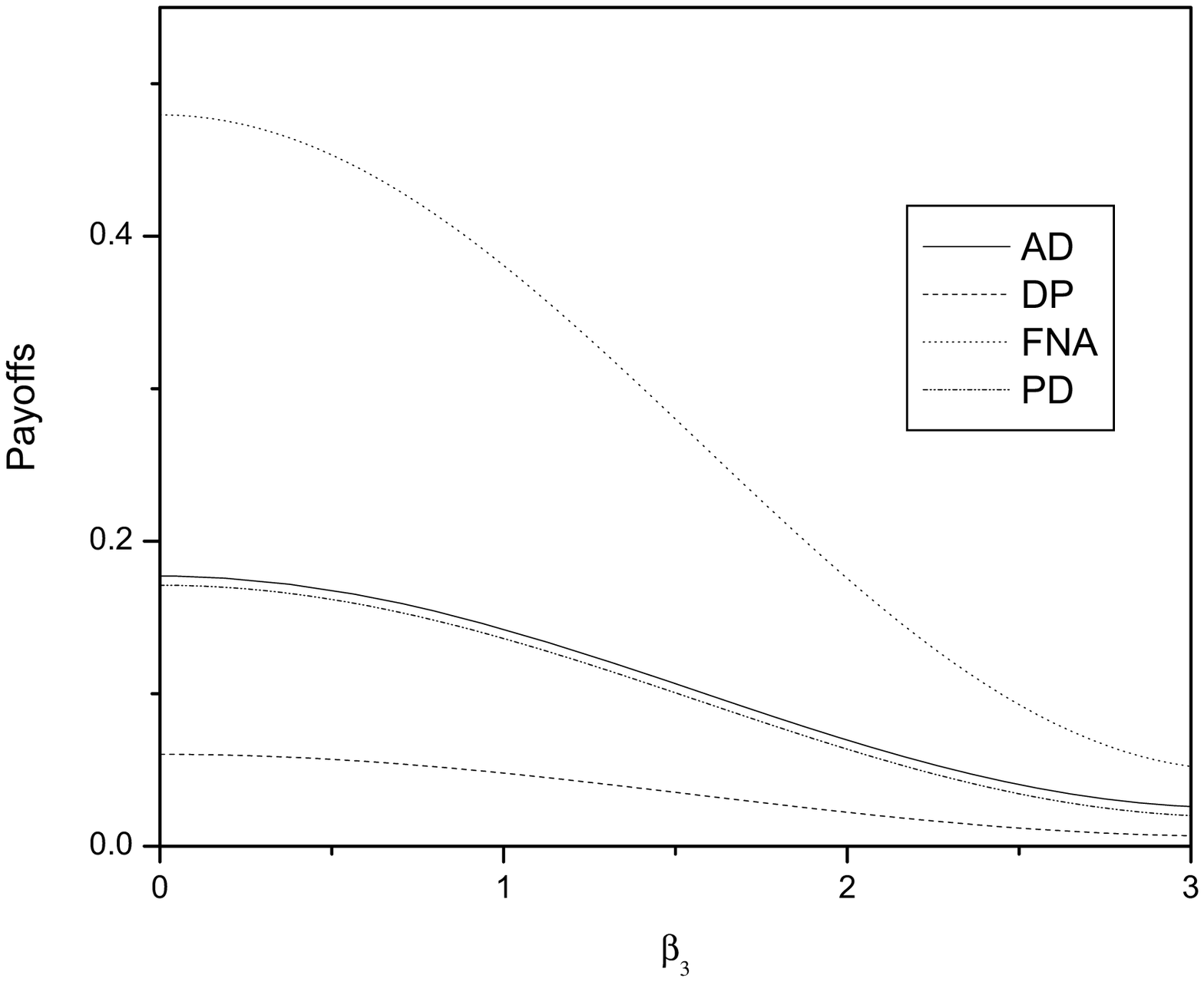} \\[0pt]
\end{center}
\caption{The expected payoffs for a single game of sequence $AAB$ are
plotted as a function of the quantum phase angle, $\protect\beta _{3}$ for
amplitude damping channel, depolarizing channel, phase damping channel and
FNA results with $p=0.5$, $\protect\delta =\frac{\protect\pi }{2}$, $\protect%
\beta _{1}=2\protect\pi $, $\protect\beta _{2}=\frac{\protect\pi }{6}$, $%
\protect\beta _{4}=\protect\pi $ and $\protect\varepsilon =\frac{1}{168}$.}
\end{figure}

\begin{figure}[tbp]
\begin{center}
\vspace{-2cm} \includegraphics[scale=0.6]{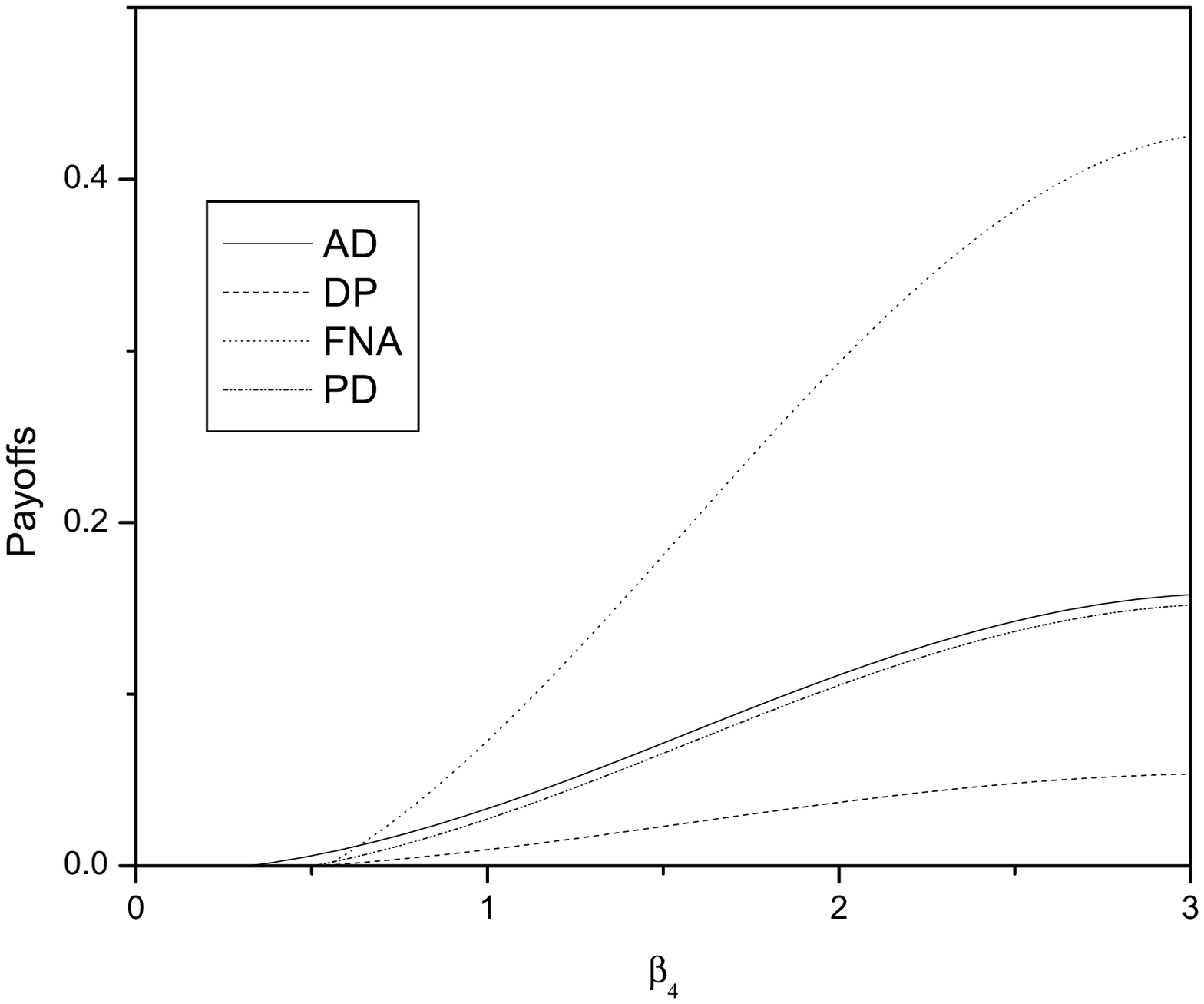} \\[0pt]
\end{center}
\caption{The expected payoffs for a single game of sequence $AAB$ are
plotted as a function of the quantum phase angle, $\protect\beta _{4}$ for
amplitude damping channel, depolarizing channel, phase damping channel and
FNA results with $p=0.5$, $\protect\delta =\frac{\protect\pi }{2}$, $\protect%
\beta _{1}=\frac{\protect\pi }{4}$, $\protect\beta _{2}=\frac{\protect\pi }{4%
}$, $\protect\beta _{3}=\frac{\protect\pi }{4}$ and $\protect\varepsilon =%
\frac{1}{168}$.}
\end{figure}
\begin{figure}[tbp]
\begin{center}
\vspace{-2cm} \includegraphics[scale=0.6]{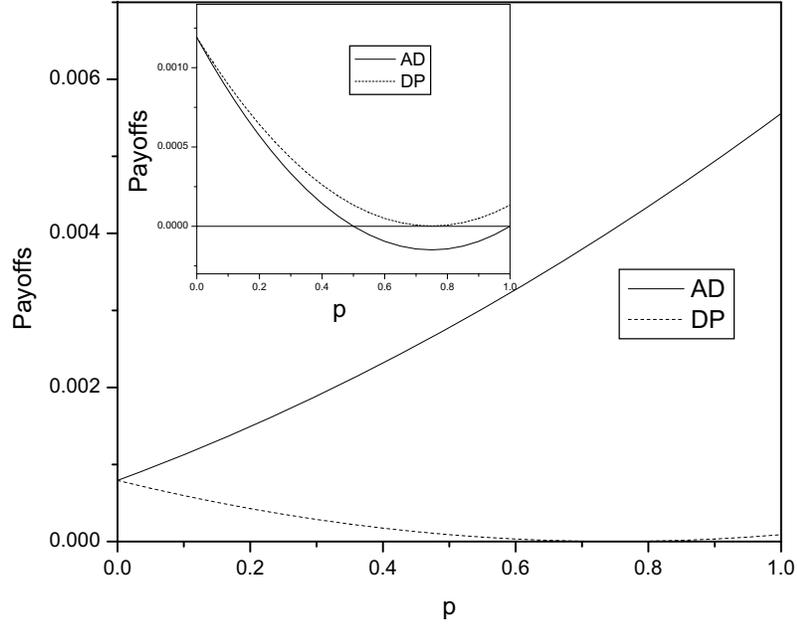} \\[0pt]
\end{center}
\caption{The expected payoffs for a series of sequence $AAB$ are plotted as
a function of the decoherence parameter, $p$ for amplitude damping and
depolarizing channels with $\protect\varepsilon =\frac{1}{168}$ (for inset
figure, $\protect\varepsilon =\frac{1}{112}$).}
\end{figure}

\begin{figure}[tbp]
\begin{center}
\vspace{-2cm} \includegraphics[scale=0.6]{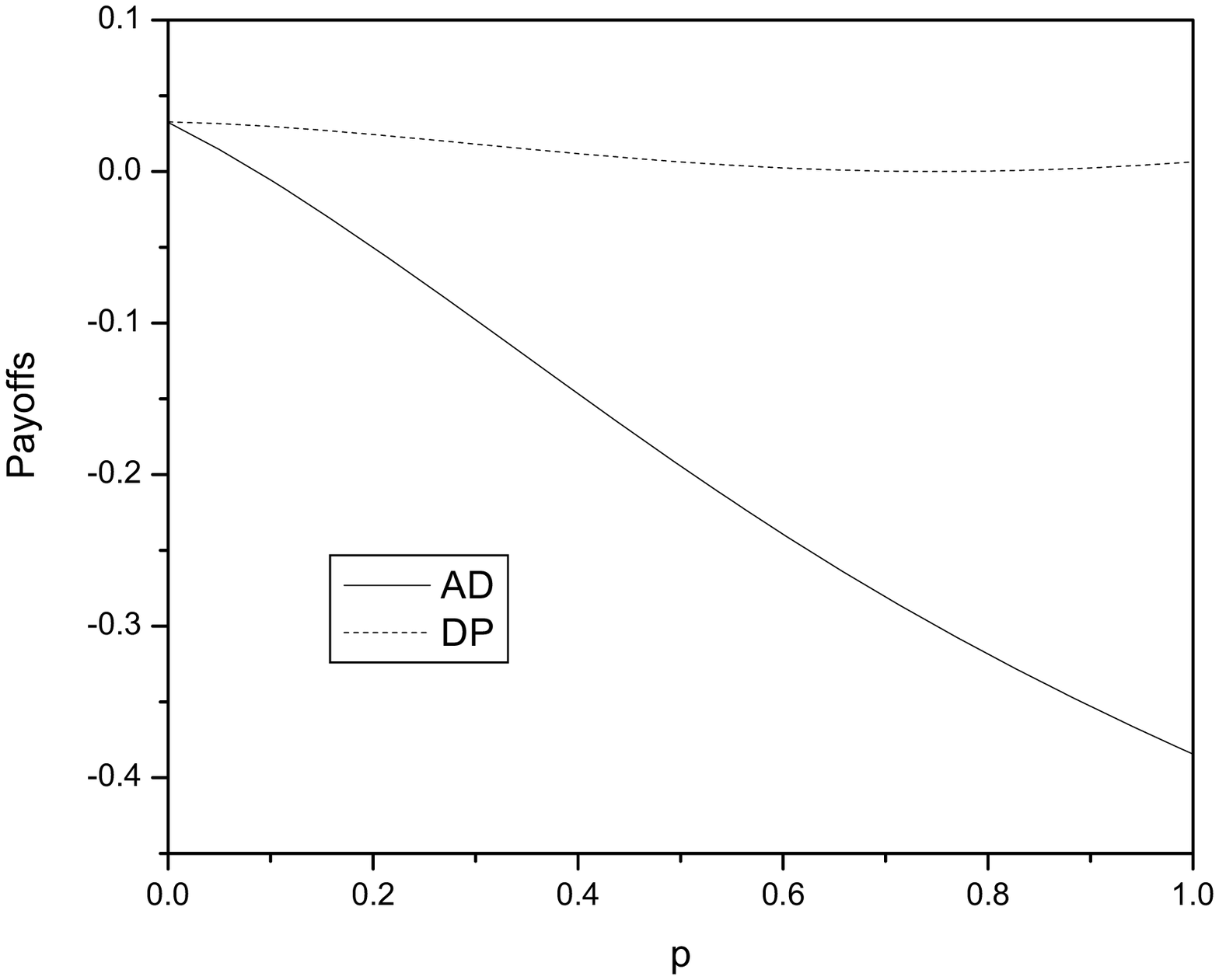} \\[0pt]
\end{center}
\caption{The expected payoffs for the game sequence $BB$ are plotted as a
function of the decoherence parameter, $p$ for amplitude damping and
depolarizing channels with $\protect\varepsilon =\frac{1}{112}$.}
\end{figure}

\begin{figure}[tbp]
\begin{center}
\vspace{-2cm} \includegraphics[scale=0.6]{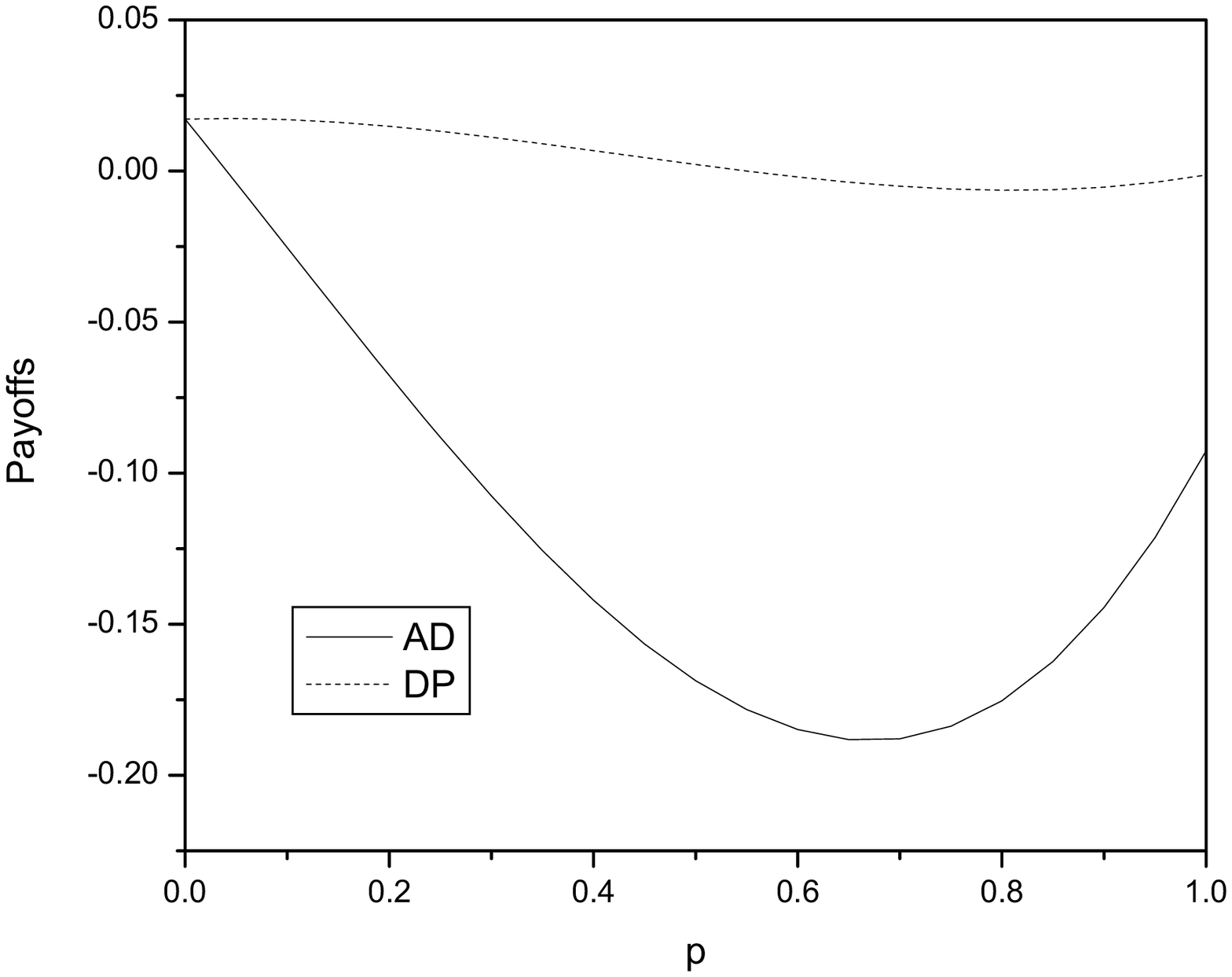} \\[0pt]
\end{center}
\caption{The expected payoffs for the game sequence $BBB$ are plotted as a
function of the decoherence parameter, $p$ for amplitude damping and
depolarizing channels with $\protect\varepsilon =\frac{1}{112}$.}
\end{figure}
\begin{table}[tbh]
\caption{Single qubit Kraus operators for typical noise channels such as
depolarizing, amplitude damping and phase damping channels where $p$
represents the decoherence parameter.}$%
\begin{tabular}{|l|l|}
\hline
&  \\
$\text{Depolarizing channel}$ & $%
\begin{tabular}{l}
$E_{0}=\sqrt{1-3p/4}I,\quad E_{1}=\sqrt{p/4}\sigma _{x}$ \\
$E_{2}=\sqrt{p/4}\sigma _{y},\quad \quad $\ $\ E_{3}=\sqrt{p/4}\sigma _{z}$%
\end{tabular}%
$ \\
&  \\ \hline
&  \\
$\text{Amplitude damping channel}$ & $E_{0}=\left[
\begin{array}{cc}
1 & 0 \\
0 & \sqrt{1-p}%
\end{array}%
\right] ,$ $E_{1}=\left[
\begin{array}{cc}
0 & \sqrt{p} \\
0 & 0%
\end{array}%
\right] $ \\
&  \\ \hline
&  \\
$\text{Phase damping channel}$ & $E_{0}=\left[
\begin{array}{cc}
1 & 0 \\
0 & \sqrt{1-p}%
\end{array}%
\right] ,E_{1}=\left[
\begin{array}{cc}
0 & 0 \\
0 & \sqrt{p}%
\end{array}%
\right] $ \\
&  \\ \hline
\end{tabular}%
$%
\end{table}

\end{document}